\documentstyle[epsf,aps]{revtex}
\begin{document}
\baselineskip=1.2\baselineskip
\title{Low-energy Pion-nucleon Scattering}

\author{
      W. R. Gibbs, Li Ai \\
 Department of Physics, New Mexico State University \\
 Las Cruces, New Mexico 88003, USA
\and
      W. B. Kaufmann \\
Department of Physics and Astronomy, 
Arizona State University\\
Tempe, Arizona 85287, USA }
\maketitle

\begin{abstract}

An analysis of low-energy charged pion-nucleon data  from recent $\pi^{\pm}p$
experiments is presented. From the scattering lengths and the GMO sum rule we
find a value of the  pion-nucleon coupling constant of $f^2=0.0764\pm 0.0007$. 
We also find, contrary to most previous analyses, that the scattering volumes
for the $P_{31}$ and $P_{13}$ partial waves are equal, within errors,
corresponding to a symmetry found in the hamiltonian of many theories.   For
the potential models used, the amplitudes are extrapolated  into the
subthreshold region to estimate the value of the $\Sigma$-term. Off-shell
amplitudes are also provided.

\end{abstract}

\newpage

\section{Introduction}

The pion-nucleon interaction has been a fruitful  source of knowledge of the
strong interaction.  The properties of the baryon  resonances produced in
pion-nucleon collisions give strong support for the quark model.  The
subthreshold amplitude is related to the value of the  pion-nucleon
$\Sigma$-term which constrains models of  nucleon structure. The pion-nucleon
coupling constant provides fundamental input for the calculation of nuclear
forces.  Beyond these surface features lie more subtle indications of the
nature of the strong interaction.

We model the dynamics of this interaction by a coupled-channel   Klein-Gordon
equation whose potential is assumed to be the fourth  component of a Lorentz
vector.  There are several advantages to this  approach. 

First, unitarity, Coulomb corrections, multichannel effects, and hadronic 
mass splittings may be included in a natural way. 

Next, by solving the Klein-Gordon equation at the appropriate kinematical
point, any observable can be calculated, even in the subthreshold region. 
This is especially easy in the case of the s-wave scattering with an 
exponential potential, for which an analytical result is available.  

By using such a model we are presented with an alternative approach to such
quantities as the  $\Sigma$-term. While the value of the $\Sigma$-term may well
be more accurately determined by dispersion relations, by looking at it from
the point of view of a potential theory the structure of the system is perhaps
better revealed in the  sense that the behavior of the subthreshold amplitude
is directly related to the shape of the potential (and presumably to the
distribution of the  constituents of the pion and nucleon).

A more subtle advantage is that  models which take a broader view of the
system, including higher energy data, have the problem that the model must be
valid over the entire range.  In this way the low-energy parameters have been
largely determined by the data at high energies and the assumed dependence of
the model for the low-energy behavior.  Thus such features as the singularities
in the scattering  amplitude due to cuts coming from the range of the
interaction depend almost entirely on the model assumptions. In the present
technique, which does not claim to be a theory of the system, the low-energy
regime can be investigated without recourse to data at higher energies.
 
Finally, solutions in coordinate space allow us to develop an intuitive picture
of the spatial structure of the interacting pion-nucleon system.  

Our Klein-Gordon model also has several drawbacks: The model is purely 
phenomenological at the  hadronic level. Because it is a model based on static
potentials,  virtual particle production and annihilation and retardation
effects are not explicitly included. As with all potential models, effects of
crossing symmetry  must be inserted. Because of the efficiency of the Jost
calculation of the s-wave amplitude for the exponential potential (and the
rapid cut-off properties of the Gaussian potential in $r$-space)  we are able
to incorporate (in a controlled approximation)  the crossing symmetry condition
that the isovector amplitude must vanish at the Cheng-Dashen (CD) point into
the fitting procedure. The relativistic effect in the center-of-mass motion is
taken into account  only approximately.  Because we describe only low-energy
phenomena, we  believe that these defects are outweighed by the model's
strengths.

In a previous paper \cite{gak} we presented results bearing on the breaking  of
isospin using the same technique applied here \cite{sg} with a selection  of
five different forms of potential.  On examining the phase shifts produced
outside the range of the fit it was noticed that the prediction was much better
for two of the models, sums of local Yukawa and exponential potentials for each
partial wave.  From fitting the data from 30 to 50 MeV these two models were
able to predict the existence of the 33 resonance, and indeed the position of
the resonance to within about 10\% in kinetic energy.  Thus it seemed
reasonable to extend the analysis to somewhat higher energy with one (or both)
of these two models.

The Yukawa potential has the advantage that it naturally represents particle
exchanges. As will be discussed in Sec. 3, the Klein-Gordon equation contains a
term which is quadratic in the potential.  For a Yukawa potential, such a term
is singular ($\propto 1/r^2$). While solutions of this equation are obtainable,
the result would seem to be more physical if a cutoff is introduced to soften
the potential at small values of $r$ (perhaps due to the intrinsic size the of
quark-pion system).  For an analysis employing particle exchange (hence
approximately related to  Yukawa potentials) see the recent work of
Timmermans\cite{timmermans}. 

One goal of this article is the representation of the pion-nucleon  interaction
in a  simple and transparent manner. The exponential potential lends itself to
that end without introducing a singular potential. Another advantage of the
exponential form is that it might better represent the interaction of the pion
with quark distributions within the nucleon. (The density calculated from a
bound-state solution of a wave equation with a $1/r$ potential is exponential,
as in the hydrogen atom and a three body system tends to follow the same
density\cite{gavin}.) 

We have also made a short study utilizing the Gaussian potential, even though
it is not considered to have a strong physical basis, to estimate which results
are sensitive to the form of the potential  (see Sec. \ref{gauss}).

The article is organized as follows. Section 2 summarizes our choice of 
elastic-scattering data sets and briefly discusses their consistency.   Section
3 reviews Jost's method for determining the s-wave amplitude for sums of
exponential potentials.  Our method of evaluation of the subthreshold
amplitudes is described in Section 4. Section 5 begins by  presenting numerical
values of the potential parameters determined from a fit to the data.  This is
followed by subsections which present the consequences of the fit, such as the
phase shifts, scattering lengths, coupling constant, sigma term, off-shell
amplitudes, partial integrated elastic cross section, and polarization
asymmetry. The results are  summarized in Section 6.

\section{Data}

The data sets that we have considered come from, for the most part, 
experiments dedicated to the measurement of pion-nucleon scattering.  We  have
excluded experiments which have not been published or which had the measurement
of pion-nucleon scattering as a secondary goal. The sets used were:

{\bf Sigg}\ \ \ The one atomic measurement by Sigg et al.\cite{sigg} is very
important in determining the low-energy behavior of the s-wave amplitudes. It
alone fixes, to large extent, the scattering length of the $\pi^-$ -p system.  
We found that predictions of this scattering length from our fits to scattering 
data were always near to their value and, when included, the fit adapted itself
easily to the value and usually fit it with a very small $\chi^2$. Thus we see
no reason to question the validity of this point and have used it in all of the
analyses discussed below.

{\bf Brack}\ \ \ The data from J. Brack et al.\cite{brack1,brack2} seem to be
of high quality.  The smallest errors are (in general) those quoted in this
work, which contains 62 data points for scattering of $\pi^+$\ and $\pi^-$\ 
at 29.4, 45.0, 66.8 and 86.8  MeV.

{\bf Frank}\ \ \ The data from J. Frank et al.\cite{frank} were the first
modern data to be published contradicting the Bertin et al.\cite{bertin}  data. 
This work contains 162 data points for scattering of $\pi^+$\ and $\pi^-$\ at 
29.4, 49.5 and 69.6 MeV. 

{\bf Auld}\ \ \  The data of E. G. Auld et al.\cite{auld} are somewhat old now
and have large error bars but are generally consistent with the modern data. 
These data contain 11 points for scattering of  $\pi^+$\ at 47.9 MeV.

{\bf Ritchie}\ \ \ The data from B. Ritchie et al.\cite{ritchie} consist of 28
points of back-angle positive pion scattering at 65.0,  72.5 and 80.0 MeV. We
have found, like Timmermans\cite{timmermans}, that the normalization seems to
be flawed in this data.  We have floated the normalization of the lowest energy
data set. (This data is from a secondary experiment.)

{\bf Wiedner}\ \ \ The data of U. Wiedner et al.\cite{wiedner}  represent the
first to appear in print from PSI.  This set consists of $\pi^+$\ and  $\pi^-$\ 
scattering at 54.3 MeV.  Our fit indicates that the cross sections for the
negative pion data are too high by about 14\%.

{\bf Joram I}\ \ \ The first paper of Joram et al.\cite{joram1} presents data
in the important Coulomb-nuclear interference region for $\pi^+$\ and  $\pi^-$\
at 32.20 and 44.60 MeV.  The set contains 80 data points and seems to be
consistent with the rest of the data base.

{\bf Joram II}\ \ \ The data in the second paper of Joram et al.\cite{joram2}
present larger angle cross sections of $\pi^+$\ and $\pi^-$\ at   32.3, 44.6
and 68.6 MeV.  The authors made single-energy fits to the data in the
experimental paper.  For the $\pi^+$\ at 32.3 MeV the best $\chi^2/N$ that they
obtained, in combination with other data at this energy, was 121/58, while for
the $\pi^+$\  at 44.6 MeV the best was 95.6/46.   It was pointed out in Ref.
\cite{hoehler5} that the scattering lengths obtained from these single-energy
fits lead to values of the $\pi$NN coupling constant outside of an acceptable
range.  We have not included these two data sets in our fit.  We also have
dropped the 53.42 degree point in the $\pi^+$\ data at 68.6 MeV which is
completely out of line with the rest of the data.  This leaves 32 data points
in this set.

\begin{figure}[htb]
\epsfysize=100mm
\epsffile{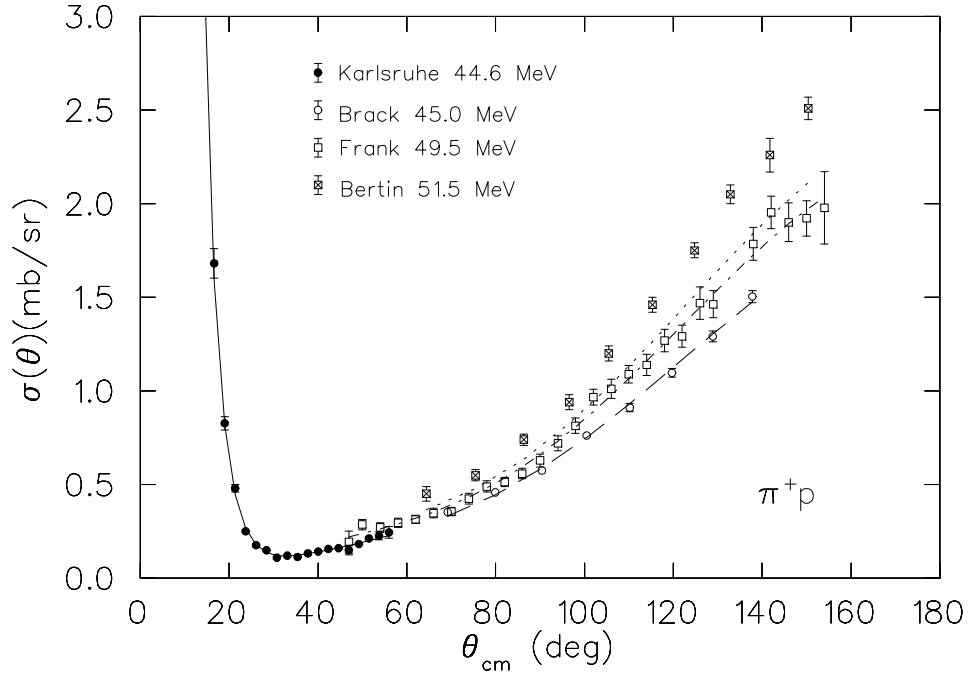}

\caption{Cross sections from $\pi^+$ proton scattering around 50 MeV.  The
Bertin data would agree with the prediction given by the dotted curve if they
were consistent with the other data sets. The solid, long dash and dash-dot
curves come from our fit and correspond to the Karlsruhe, Brack and Frank data
sets.}\label{cross} \end{figure}

The data of Bertin et al.\cite{bertin} were not included in the general fits
since they (or at least their normalizations)  seem to be  inconsistent with
the modern data.  In order to make contact with previous analyses and get some
feeling for the impact of the data sets,  we make fits using the Bertin data as
the only $\pi^+$\ data and the full modern $\pi^-$\ data set described above. 
Figure 1 shows the $\pi^+$\ data around 50 MeV  compared with one of our fits
illustrating the apparent discrepancy.

\section{Solutions for the Amplitudes: Jost Representation}\label{josts}

At low energies and for each partial-wave, the $\pi^+$ p elastic process is
described by a single-channel Klein-Gordon (KG) equation.  The $\pi^-$ p 
elastic
and charge-exchange scattering are described by a two-channel KG equation. 
That the effect of the ($\pi^-,\gamma$) reaction  on the hadronic channels may
be ignored was justified in Ref. \cite{sg} where this procedure is discussed in
detail.

In fitting the data, we solve the KG equation by standard  numerical
procedures. The potential, $V$, is included in the KG equation through the
substitution $\omega \rightarrow \omega-V$, where $\omega$ is the center of
mass energy of the pion (actually the reduced energy in the fits); i.e.  for
both electrostatic and strong interactions, $V$ is taken to be the fourth
component of a Lorentz four-vector. The resulting equation is

\begin{equation} 
(\nabla^2+k^2-2\omega V+V^2)\psi =0.
\end{equation}
where $k$ is the center-of-mass momentum.

However there are several calculations that we would like to make which include
only the strong interaction (the sigma term, the isovector s-wave amplitude at
the CD point, and the off-shell amplitudes)  for which the Coulomb force is not
included.  In this case it is very convenient to use (for the s-wave) the
expressions developed by Jost\cite{jost}. 

We now give a brief outline of the method and its extension to the off-shell
case. Our potential contains (at most) two terms of exponential form.  Since
the potential appears linearly and quadratically in the KG equation, a
two-term basic potential leads to an effective Schr\"odinger equation with a
5-term potential. 

Consider the solution of the Schr\"odinger equation for a sum of exponential
potentials:

\begin{equation} 
V(r)=\sum_{j=1}^N \lambda_je^{-\mu_jr}.
\end{equation}
Jost \cite{jost} writes the solution for the s-wave, $f(k,r)$, as

\begin{equation} 
f(k,r)=e^{-ikr}\sum_{\alpha}C_{\alpha}(k)e^{-m_{\alpha}r} 
\end{equation}
where the subscript $\alpha$ is a compound quantity consisting of a
set of N integers.  For example, for a three-term potential

\begin{equation} 
\alpha\equiv [j,k,l],\ \ j,k,l=0,1,2 \dots 
\end{equation}
The coefficients $C_{\alpha}(k)$ are given by the recursion relation
\begin{equation} 
C_{[j,k,l]}(k)=\frac{\lambda_1C_{[j-1,k,l]}+\lambda_2C_{[j,k-1,l]}
+\lambda_3C_{[j,k,l-1]}}{m_{\alpha}(m_{\alpha}+2ik)}\label{recur} 
\end{equation}
where
\begin{equation} 
m_{\alpha}\equiv m_{[j,k,l]}\equiv j\mu_1+k\mu_2+l\mu_3 .
\end{equation}
The recursion is started with
\begin{equation} 
C_{[0,0,0]}=1, \ \ C_{[-1,k,l]}=C_{[j,-1,l]}=C_{[j,k,-1]}=0
\end{equation}
and is built up by first computing all coefficients with the sum of
indices equal to one, then two, {\it etc.} with no negative index.

The solution with the proper boundary conditions at the origin with an incoming
spherical wave and with unit amplitude at infinity is

\begin{equation} 
\psi(k,r)=-\frac{f(k,r)-S(k)f(-k,r)}{2ikr} \label{psis}, 
\end{equation}
where
\begin{equation} 
S(k)=\frac{f(k,0)}{f(-k,0)}. 
\end{equation}
These expressions can be used to calculate the values of the S-matrix for any
value of $k$.  We shall be interested in  purely imaginary values for the
calculation of the s-wave  contribution to the $\Sigma$-term.

In order to calculate the off-shell amplitude we consider the wave 
function for real (positive) values of $k$.  In this
case we can write (for real $\lambda_j$)
\begin{equation} 
f(-k,r)=f^*(k,r)\ \ {\rm with}\ \ S=e^{2i\delta}\ \ {\rm and}\ \ 
 f(k,0)=e^{i\delta}\beta 
\end{equation}
where $\beta$ is real and positive.  We can now write the wave function (Eq. 
\ref{psis}) as
\begin{equation} 
\psi(k,r)=-\frac{e^{i\delta}}{2ikr}\left[
e^{-i\delta}e^{-ikr}\sum_{\alpha}C_{\alpha}(k)e^{-m_{\alpha}r}-
e^{i\delta}e^{ikr}\sum_{\alpha}C_{\alpha}(-k)e^{-m_{\alpha}r}\right] 
.\end{equation}
The s-wave off-shell amplitude is defined by
\begin{equation} 
F_0(k,q)\equiv\int dr r^2 j_0(qr)V(r)\psi(k,r)
\end{equation}
and we find
\begin{equation} 
F_0(k,q)=\frac{e^{i\delta}}{k}Im 
\left[e^{-i\delta}\sum_{\alpha}C_{\alpha}(k)
\sum_j\frac{\lambda_j}{(m_{\alpha}+\mu_j)^2+q^2-k^2+
2ik(m_{\alpha}+\mu_j)}\right]. 
\end{equation}
This function is evaluated in Sec. \ref{osa}.

\section{Subthreshold Extrapolation}

The subthreshold regime, the region below the elastic threshold, is of 
interest in studies of chiral-symmetry breaking, a measure of which is 
the $\Sigma$ term. On-shell subthreshold amplitudes have been evaluated 
over the years by the use of dispersion relations 
\cite{hoehler,campbell,koch,hite,sainio}. It is interesting to compare 
those results with the values given by the KG equation.

\subsection{Dispersion Relations}

Because the ``experimental'' $\Sigma $-term is defined at the unphysical CD
point ($s = m^2$, $t = 2\mu^2$), any determination of it is, to a degree, model
or theory dependent. $s$-channel and $u$-channel data exists only for negative
$t$, therefore fixed-$t$ dispersion relations must either rely on the rapid
convergence of the partial-wave series outside of the physical region at $t=2
\mu^2$ as in Cheng and Dashen's method or must use additional techniques (such
as dispersion relations at fixed $\nu$) to extend the amplitudes to $t=2
\mu^2$.  This extrapolation is complicated by the presence of the crossed
reaction $\pi \pi \rightarrow N\overline{N}$ which produces a nearby branch cut
in the t-channel beginning at $t=4\mu ^2$. Some knowledge of the effect of this
cut is obtained indirectly from $\pi \pi $ elastic scattering, but this
analysis is, to an extent, model dependent. Some formulations of the dispersion
relation approach rely heavily on very accurate knowledge of the p-wave
scattering volume\cite{sainio} when, in fact, the $\Sigma$ term is dominated by
the s-wave amplitude\cite{ericson,hoehler2}. 

Interior Dispersion Relations (IDR) \cite{hite} and Hyperbolic Dispersion
Relations \cite{koch} have also been used to determine the $\Sigma $-term.  In
these cases, the curves along which the dispersion relations are written may be
chosen to pass directly through the CD point. However, these dispersion
relations involve integrals over the entire $t$ -channel cut, a long portion of
which is unphysical ($4\mu ^2\leq t<4m^2$). In this case, the entire
$t$-channel dispersion integral (the ``Discrepancy Function'') must be
determined and extrapolated to the CD point. Thus, even in the classic
dispersion-theory determination of the $\Sigma $-term, some model dependence is
present.

\begin{figure}[htb]
\epsfysize=100mm
\epsffile{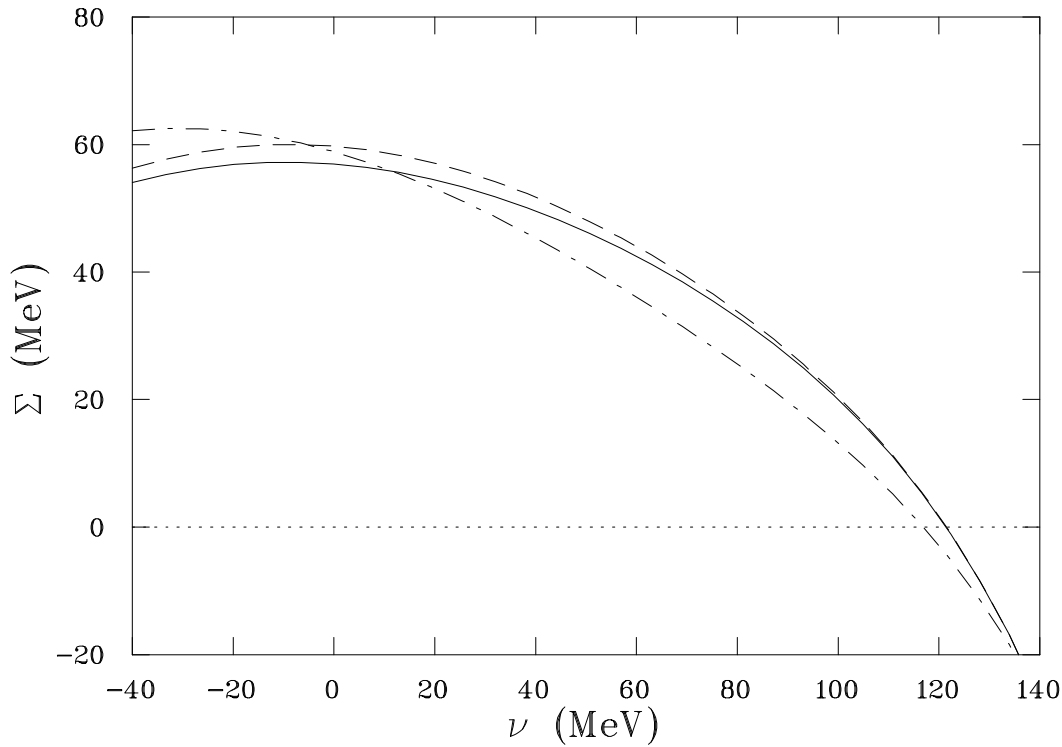}
\caption{Contributions to the $\Sigma$ term for the s-wave (dash-dot
curve) the s- and p-waves (dashed curve) and the full amplitude 
(solid curve) from the polynomial expansion from Ref. 
\protect\cite{hoehler4} From this graph the p-wave is estimated to give
a correction of $\sim$ +1 MeV. The sum of all of the non-s-wave contributions 
gives $\sim$ -1.5 MeV.}\label{sigmaps} \end{figure}
 
It is instructive to compare the relative contributions of the $s$ and $p$
waves to the non-flip pole-subtracted amplitude $\tilde{G}^{+}$ which is real
in this region and, as we shall show in the next section, is proportional to
$\Sigma$ at the CD point. To this end we have performed the partial-wave
projections of  $\tilde{G}^{+}$ as computed from the subthreshold polynomial
expansion in $\nu =(s-u)/4m$ and  $t$ from H\"ohler et al.\cite{hoehler4}.  For
purposes of illustration we plot $\tilde{G}^{+}$ {\it vs.} $\nu$  along a curve
of constant path parameter $a=-\phi (s,t)/t^2,$ where $\phi (s,t)=t\left[
su-(m^2-\mu^2)^2\right]$ is the Kibble function, which vanishes on the boundary
of the physical region.  The parameter, $a$, is fixed at the value
$-m^2+\mu^2/2$.  This path has the advantage that it passes through both the 
$s$-channel threshold ($\nu =\mu $) and the CD point (at which $\nu =0$). It
also arrives at the CD  point tangent to a curve of constant $t$ as was
specified in the paper of Cheng and Dashen. (This is the path used in IDR
calculations.) For comparison, the $s$ and  $s+p$ wave partial-wave
contributions to $\tilde{G}^{+}$  are plotted along with the full amplitude in
Fig. \ref{sigmaps}. The amplitudes have been multiplied by a factor so that
they reduce to the $\Sigma $-term at $\nu =0$. It is seen that the $s$ wave
contribution is  similar to the full amplitude $\tilde{G}^{+}$ between the CD
point and threshold, becoming quite close at the CD point. Notice that the $s$
and $s+p$ contributions cross slightly below the CD point, where $\cos \theta $
vanishes. This zero in $\cos \theta $ suppresses the $p$-wave contribution at
the CD point\cite{ericson}.  The value of $\cos \theta$ at the CD point is

\begin{equation} 
\cos\theta=-\frac{\mu^2}{4m^2-\mu^2} \approx -\frac{1}{180}. 
\end{equation} 
The value of $t$ for which $\cos\theta$ is zero (for $\nu =0$) is approximately 
\begin{equation} 
t=2\mu^2(1-\frac{\mu^2}{4m^2}) 
\end{equation} 
which differs from the value at the CD point by about 1/2 of one percent. 

\subsection{Klein-Gordon Approach}

In the present approach the subthreshold dependence  is derived directly from
the pion-nucleon data itself. The question of the validity of  the function
extracted then depends on the accuracy of the data and the suitability of the
form used for the potential.  

While the amplitudes produced from the KG model do not automatically contain
the full analyticity and crossing properties of the invariant amplitudes, much
of the same physics can be included by  considering that the source of the
potential is the $t$-channel cut ($\pi\pi$ scattering and $\pi\pi\rightarrow
N\bar{N}$). The potential can be written in terms of an integral over the
discontinuity along the $t$ axis as done in the case (for example) of the $NN$
potential\cite{paris}

\begin{equation} 
V(r)=\int_{4\mu^2}^{\infty} \rho(t)\frac{e^{-\sqrt{t}r}}{r}dt.
\end{equation}
Since a Yukawa potential can be written as an integral over 
an exponential form
\begin{equation} 
\frac{e^{-\mu r}}{r}=\int_{\mu}^{\infty}e^{-xr}dx 
\end{equation}
the $\pi N$ potential can be expressed as a sum of exponential potentials.

From the definition of the non-spin-flip amplitude,
\begin{equation}
G(s,t)=\sum [(2\ell +1)f_{\ell +}+\ell f_{\ell -}]P_{\ell}(x),
\end{equation}
and the projection of these partial waves from the invariant amplitudes
$A(s,t)$ and $B(s,t)$ (see Ref. \cite{nielsen} Eq. 2.7 for example)
we can write
$$16\pi sG^{(+)}(s,t)=[(W+m)^2-\mu^2][A^{(+)}(s,t)+(W-m)B^{(+)}(s,t)]$$
\begin{equation}
+x[(W-m)^2-\mu^2][-A^{(+)}(s,t)+(W+m)B^{(+)}(s,t)] 
\end{equation}
where $W=\sqrt{s}$ and $x$  ($=\cos\theta$) is a function of $s$ and 
$t.$ The same result is available from Eq. A 2.32 of Ref. \cite{hoehler}.
We omit the superscript ``(+)'' for the next few equations for clarity.
For $t=2\mu^2$ the Born term contribution to $B$,
$ g^2\left(\frac{1}{m^2-s}-\frac{1}{m^2-u}\right)$,
becomes $-\frac{2g^2}{s-m^2}$.

\newpage

If we define $\tilde{B}(s,t)$ as the $B$-amplitude with the pole term
removed then

$$ 
16\pi sG(s,2\mu^2)=
\left[(W+m)^2-\mu^2\right]\left[A(s,2\mu^2)+(W-m)\tilde{B}(s,2\mu^2)
-\frac{2g^2}{W+m}\right]$$
\begin{equation}
+x\left[(W-m)^2-\mu^2\right]\left[-A(s,2\mu^2)+(W+m)\tilde{B}(s,2\mu^2)
-\frac{2g^2}{W-m}\right] 
\end{equation}
or

$$
\left[(W+m)^2-\mu^2\right]
\left[A(s,2\mu^2)+(W-m)\tilde{B}(s,2\mu^2)-\frac{2g^2}{W+m}\right]
=16\pi sG(s,2\mu^2)-\frac{2g^2\mu^2x}{W-m} 
$$
\begin{equation}
-x(W-m)^2[-A(s,2\mu^2)+(W+m)B(s,2\mu^2)]
+x\mu^2[-A(s,2\mu^2)+(W+m)\tilde{B}(s,2\mu^2)]
\end{equation}
Now define
$\tilde{G}(s,2\mu^2)\equiv G(s,2\mu^2)-\frac{\mu^2g^2x}{8\pi s(W-m)}$ to 
remove the pole from the $P_{11}$ partial wave.
The subtracted pole term should be evaluated with the fitted position
and residue so that the pole is exactly removed from the $P_{11}$ partial
wave of $G(s,2\mu^2)$. Then, at the CD point ($s=m^2$),
\begin{equation}
(4m^2-\mu^2)\left[A(m^2,2\mu^2)-\frac{g^2}{m}\right]
=16\pi m^2\tilde{G}(m^2,2\mu^2)
+x_{CD}\mu^2\left[-A(m^2,2\mu^2)+2m\tilde{B}(m^2,2\mu^2)\right].\label{eq22}
\end{equation}

The last term is negligible at the CD point as we will show shortly.  
Thus, neglecting the factor $(1-\frac{\mu^2}{4m^2})$,
\begin{equation}
 4\pi\tilde{G}^{(+)}(m^2,2\mu^2)=A^{(+)}(m^2,2\mu^2)-
\frac{g^2}{m}=\Sigma/f_{\pi}^2
\end{equation}
where $f_{\pi}= 93.2 $ MeV is the pion decay constant. 
The equation corresponding to the second equal sign has been given many places 
(see \cite{no} for example). The corrections to this expression of the order
$(\frac{\mu}{m})^4$ \cite{brown}.

The ratio of the last two terms in Eq. \ref{eq22} is approximately
$$ x_{CD} \mu^2 [-A+2m \tilde{B}]f_{\pi}^2/(4 m^2 \Sigma) \approx 0.002 $$
where we have used $A \approx g^2/m \approx 191$ GeV$^{-1}$, $2 m \tilde{B}
\approx -311$ GeV$^{-1}$, $\Sigma \approx 0.060$ GeV, and $x_{CD} \approx
-\mu^2/4 m^2$.  Thus we may safely ignore the last term in Eq. (22). 

Thus, to a good approximation, the sigma term can then be evaluated as 
$\Sigma = 4 \pi  f_{\pi}^2 \tilde{G}^{(+)}$ where $\tilde{G}^{(+)}$ is 
evaluated at the CD point.  We approximate $\tilde{G}^{(+)}$ in terms of 
s-channel isospin amplitudes:
$G^{(+)}=\frac{1}{3}(G^{\frac{1}{2}}+2 G^{\frac{3}{2}})$. 

The CD point occurs at center-of-mass total pion energy,  $\omega =
\frac{\mu^2}{2m} \approx 0$; at this point, the effective potential $U=2 \omega
V-V^2 \approx -V^2$ is attractive whatever the sign of $V$. Above threshold the
cross term dominates, and $U$ may be either positive or negative. We are
particularly interested in the s-wave scattering amplitudes because,  as
discussed in the previous section, at the CD point $\cos\theta \approx 0$ and
the p-wave contributions to the sigma term are suppressed. At threshold one of
the s-wave amplitudes, $S_{1/2}$ is positive, and the other $S_{3/2}$ is
negative. We define renormalized amplitudes by 

\begin{equation} 
\Sigma_I = 4\pi f_{\pi}^2 G^{(+)}_I \approx 4\pi f_{\pi}^2 S_I
 \approx 553\ S_I \label{sigdef} 
\end{equation}
 where $S_I$ are the
s-wave scattering amplitudes in fm. and $I=\frac{1}{2},\frac{3}{2}$. When
evaluated at the CD point $\Sigma=\frac{1}{3}(\Sigma_{\frac{1}{2}}+2
\Sigma_{\frac{3}{2}})$ is the sigma term. 

As $\omega$ passes below threshold  ($\omega =\mu$) these two amplitudes will
keep their signs (they are real in this region) but at $\omega=0$ they must be
of the same sign (and positive) because the square term now dominates (unless
the squared potential is strong enough that there is a bound state). 

In the low-energy physical region there is a correlation between the  range and
strength of each potential; i.e. two fits are, to  lowest order, comparable if
the volume integrals of the potentials are equal.  However, in the subthreshold
region near the CD point the $V^2$ term dominates $2 \omega V$ ($\omega \approx
0$); the correlation no longer holds.  Therefore to determine the amplitude
near the CD point with our model, the strength and range of the potentials must
be independently determined. We believe that the available data is sufficiently
accurate that this determination can be made, at least approximately.

Since $A^{(-)}(m^2,2\mu^2)=0$ and the Born term is zero at the CD point, we
find
\begin{equation}
G^{(-)}(m^2,2\mu^2)\approx 0, 
\end{equation}
where the zero is of order $(\frac{\mu^2}{4m^2})^2$ times a typical amplitude.
Since the isovector amplitude $G^{(-)}(m^2,2\mu^2)$ vanishes,  it follows that
the isovector  combination  $\Sigma^- =\Sigma_{\frac{1}{2}} -
\Sigma_{\frac{3}{2}}$  should vanish  (assuming s-wave dominance) at the CD
point so that the two amplitudes defined in Eq. \ref{sigdef}  should cross at
that point. This vanishing is not automatic in the KG model, which doesn't have
$s \leftrightarrow u$ crossing built in.  In performing a number of fits
without this  requirement we found that $\Sigma_{\frac{1}{2}}$ and
$\Sigma_{\frac{3}{2}}$  do indeed naturally cross near $\nu =0$. However,  the
isovector amplitude varied between 10 and 100 MeV at the CD point. Even so, we
obtained stable values for the scattering lengths while  values of the $\Sigma$
term varied from 33 to 100 MeV. 

Using the Jost technique described in Section \ref{josts} we evaluated
$\Sigma^-$ at the CD point at each step of the  fit. By including a
contribution of

\begin{equation} 
\left(\frac{\Sigma_{\frac{3}{2}}-\Sigma_{\frac{1}{2}}}{\Delta 
E_{\Sigma}}\right)^2 \label{chisig} 
\end{equation} 
to the $\chi^2$ it was possible to force the vanishing of $\Sigma^-(\nu=0)$ to
a reasonable  degree of accuracy. The result is that the value of the $\Sigma$
term obtained is much more stable. See Ref. \cite{don} for a related
discussion of subthreshold constraints on potential models.  

\begin{table}[htb]
\begin{tabular}{lcrcrrr}
{\rm Partial\ Wave}& $\alpha_1$\ {\rm (MeV/c)}&$\lambda_1$ \ {\rm (MeV)}&
$\alpha_2$\ {\rm (MeV/c)}&$\lambda_2$\ {\rm (MeV)}&$R_1$\ {\rm (fm)}&
$R_2$\ {\rm (fm)} \\
${\rm S}_{3}$&723.3&901.0&485.3&-41.4&0.945&1.408\\
${\rm S}_{1}$&750.8&-1.6&310.5&-58.0&0.910&2.201\\
${\rm P}_{33}$&891.7&399.2&486.7&-1548.6&0.767&1.404\\
${\rm P}_{13}$&813.9&-5584.2&528.5&834.3&0.840&1.293\\
${\rm P}_{31}$&631.4&1800.5&&&1.083&\\
${\rm P}_{11}$&720.3&-3540.3&&&0.949&\\
\end{tabular}
\caption{Ranges and strengths for the potential obtained from a fit to the
data.  The radii are calculated in the rms sense.} \label{ranges} \end{table}

\begin{figure}[htb]
\epsfysize=100mm
\epsffile{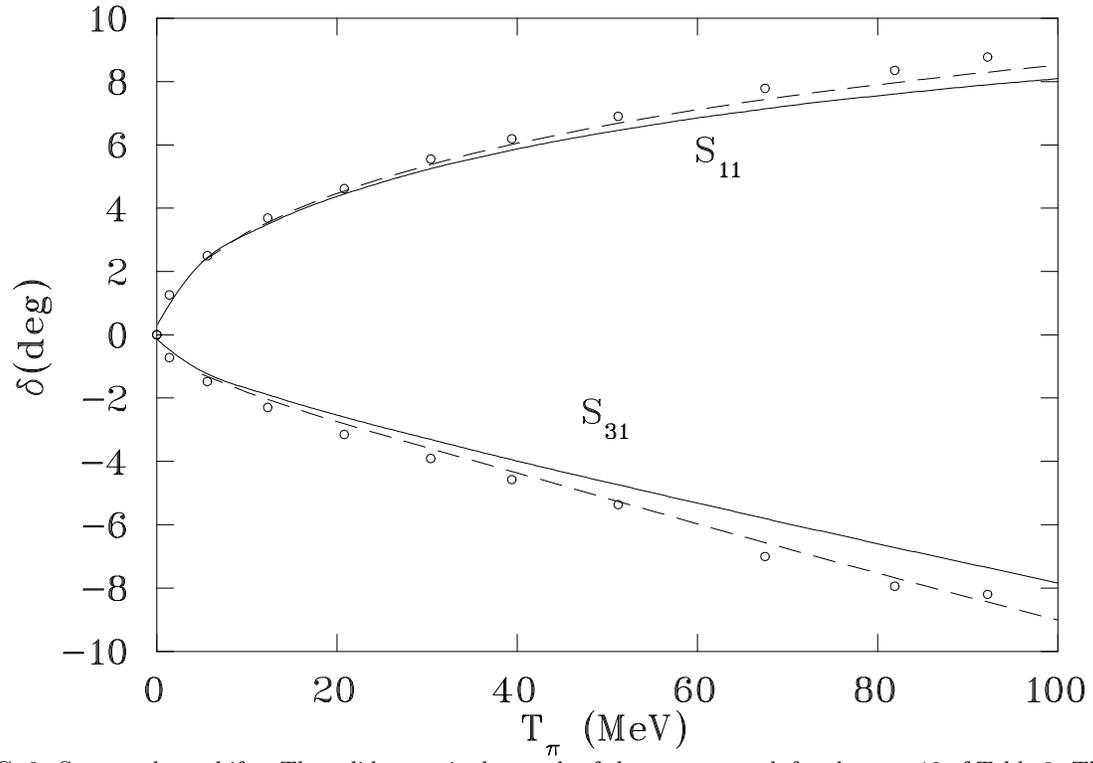}
\caption{S-wave phase shifts.  The solid curve is the result of the present
work for the case 12 of Table 3. The circles are the results of the
KH80 solutions and the dashed line is from SM95.}\label{swaves} \end{figure}

\begin{figure}[htb]
\epsfysize=100mm
\epsffile{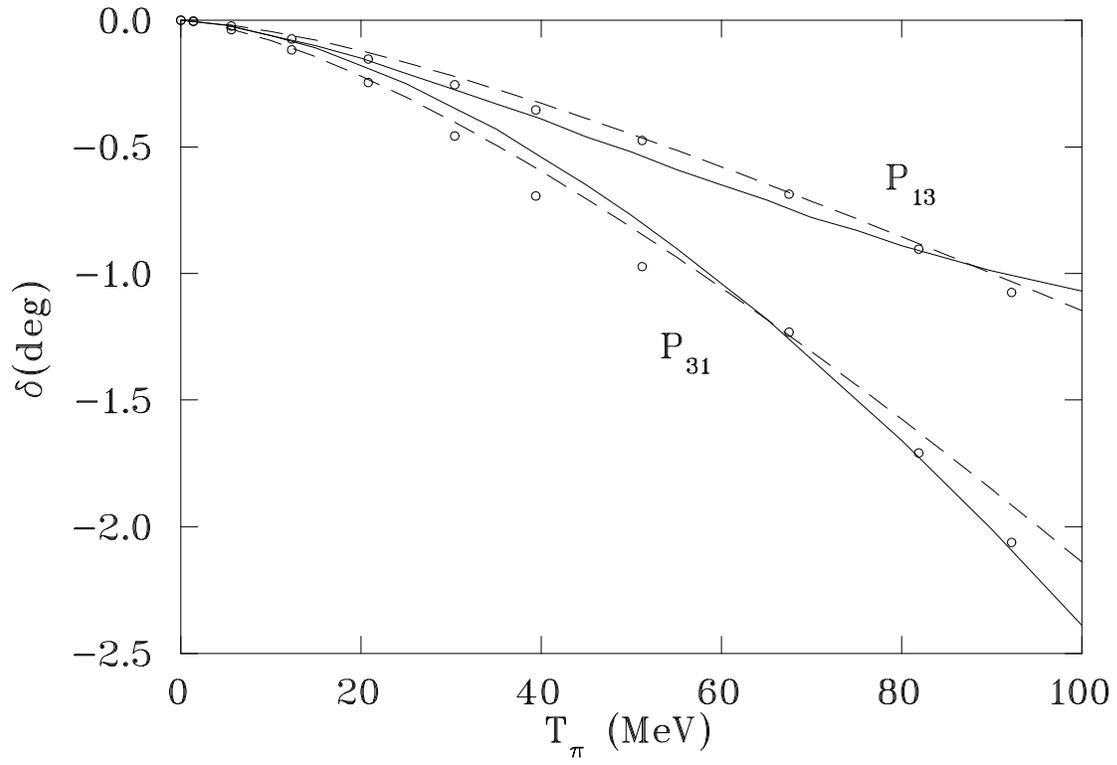}
\caption{P$_{31}$ and P$_{13}$ wave phase shifts. 
The curves and symbols have the same meaning as in Fig. 
\protect\ref{swaves}
}\label{p31p13} \end{figure}

\begin{figure}[htb]
\epsfysize=100mm
\epsffile{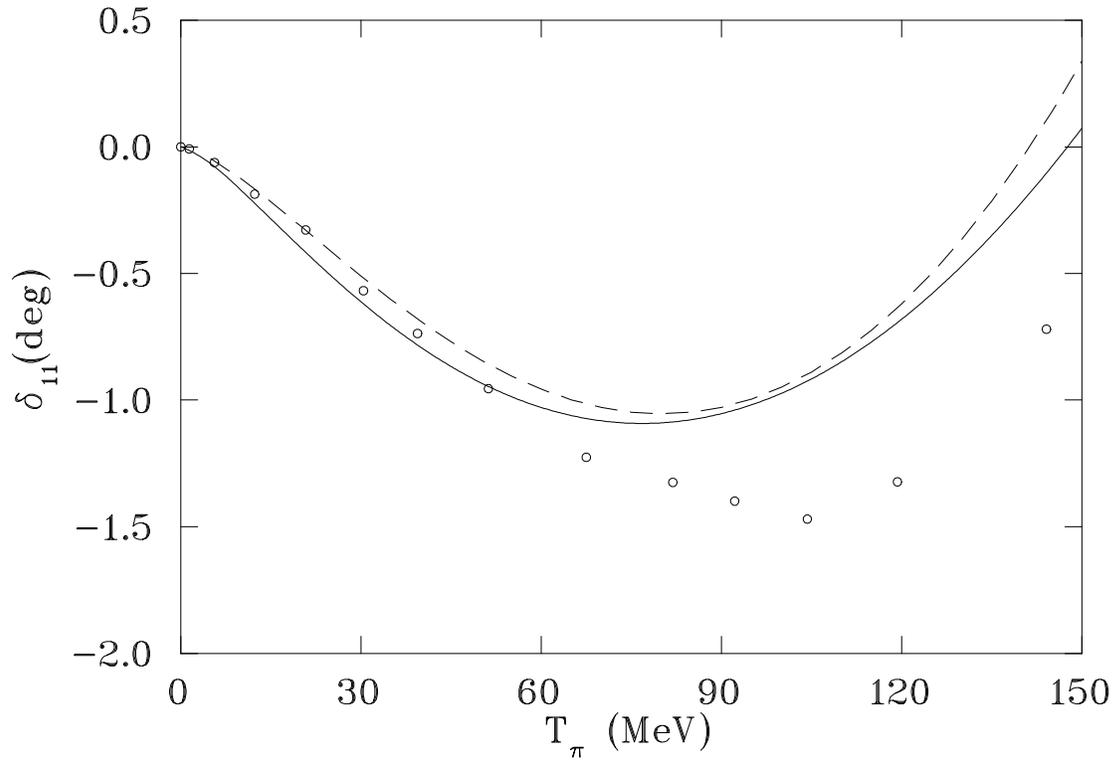}
\caption{P$_{11}$ wave phase shifts.
The curves and symbols have the same meaning as in Fig. 
\protect\ref{swaves}
}\label{p11} \end{figure}

\begin{figure}[htb]
\epsfysize=100mm
\epsffile{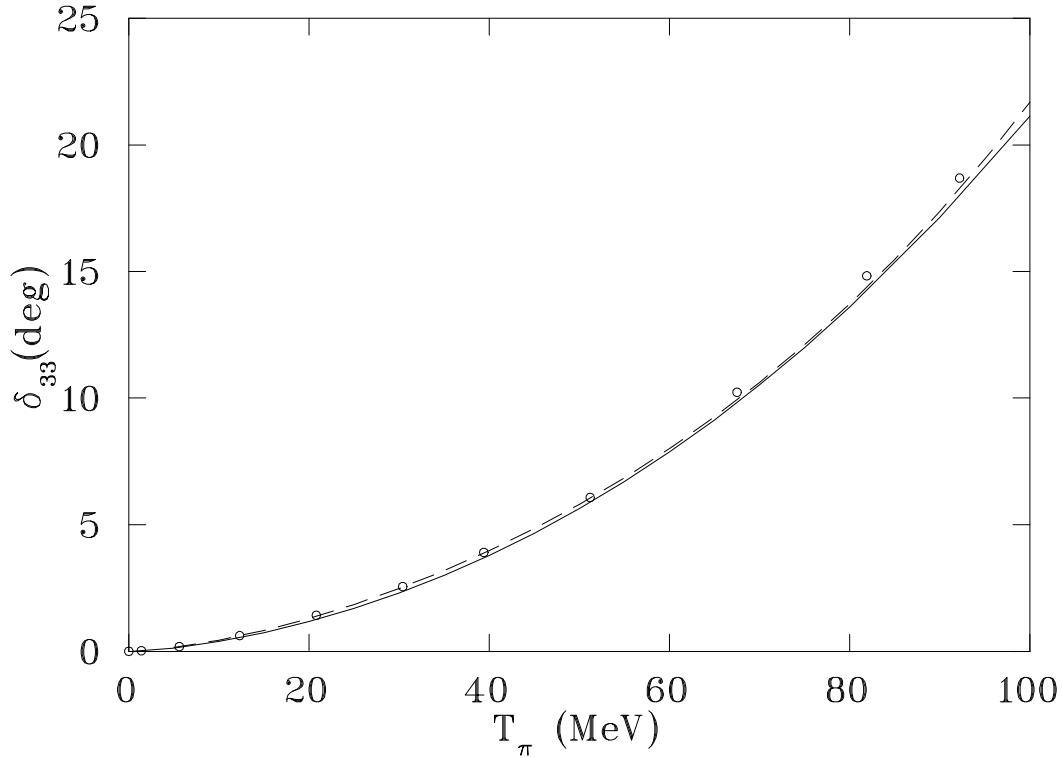}
\caption{P$_{33}$ wave phase shifts
The curves and symbols have the same meaning as in Fig. \protect\ref{swaves}
}\label{p33} \end{figure}

\section{Results of the Fit}

The form of the potential for each partial wave was taken as

\begin{equation} 
V(r)=\lambda_1e^{-\alpha_1r}+\lambda_2e^{-\alpha_2r}. 
\end{equation}
The second term was not needed for two of the partial waves. As an example of
the values obtained, the strengths and ranges are given in 
Table \ref{ranges} for the case 11 in Table 2.
We see that the ranges corresponding to the dominant strengths mostly have
values from 500-900 MeV/c.  An exception is the partial wave $S_{11}$ where
the value is $\sim 310 $ MeV/c, barely above 2 pion masses, the minimum
acceptable value without creating an anomalous threshold.  That such a
long (spatial) range is needed to fit this partial wave was noted by
Ref. \cite{jennings2}.

We have tested individual data sets to see their influence on the fits. 
The results presented in Tables \ref{results1} and \ref{results2} show 
the effect of adding, one by one, the data sets in the order shown. Below 
the solid lines are fits to single groups of data. The second line that 
appears in most entries of Table \ref{results2}  corresponds to a 
second fit which gives an idea of the uniqueness of the  fits.

\begin{table}[htb]
\begin{tabular}{lrrrrrr}
&$\chi^2$/{\rm n\ (data)}&$\chi^2$/{\rm n\ (norm)}&
 $\Sigma$ {\rm (MeV)}&$a_3$\ ($\mu^{-1}$)&$a_1$\ ($\mu^{-1}$) &$f^2$\ \ \\
{\rm Brack }&64.66/62\ &9.88/10\ \ &42.86\ &-0.083&0.174&0.0759 \\
{\rm +Frank }&231.78/228&21.85/16\ \ &48.43\ &-0.085&0.174&0.0762 \\
{\rm +Auld}&243.28/239&23.20/17\ \ &48.59\ &-0.085&0.174&0.0762 \\
{\rm +Ritchie}&274.58/267&27.68/20\ \ &48.66\ &-0.085&0.174&0.0762 \\   
{\rm +Wiedner}&316.27/306&33.09/22\ \ &50.01\ &-0.085&0.174&0.0762 \\
{\rm +Joram\ I}&432.67/386&35.42/26\ \ &48.90\ &-0.081&0.172&0.0751 \\
{\rm +Joram\ II}&497.93/417&42.01/30\ \ &53.59\ &-0.082&0.172&0.0753 \\
\hline
{\rm Frank}&144.15/166&4.56/6\ \ \ &48.34\ &-0.083&0.173&0.0757 \\
{\rm PSI}&153.49/150&8.33/10\ \ &67.45\ &-0.095&0.180&0.0793\\
\hline
{\rm Bertin}+$\pi^-$&278.56/243&15.96/18\ \ & 52.92\ &-0.105&0.183&0.0817\\
\end{tabular}

\caption{Results of the study.  $\chi^2$ is separated into contributions from
the individual data points and from the experimental normalization 
uncertainty. The fits were made including a contribution to the total 
$\chi^2$ from Eq. \protect\ref{chisig}  of  $\Delta E_{\Sigma}=10$ MeV. 
The line labelled PSI includes both the Wiedner and Joram data.} 
\label{results1}\end{table}

\begin{table}[htb]
\begin{tabular}{rlrrrrrrr}
&&$\chi^2$/{\rm N\ (Data)}&$\chi^2$/{\rm N\ (Norm)}& $\Sigma$ {\rm (MeV)}
&$a_3$\ ($\mu^{-1}$)&$a_1$\ ($\mu^{-1}$) &$f^2$\ \ \\
1&{\rm Brack }&66.37/62\ &9.70/10\ \ &50.51\ &-0.082&0.174&0.0757 \\
2&&65.09/62\ &10.52/10\ \ &45.53\ &-0.083&0.174&0.0759 \\
3&{\rm +Frank }&234.69/228&21.56/16\ \ &50.77\ &-0.085&0.175&0.0764 \\
4&&230.63/228&22.63/16\ \ &46.32\ &-0.086&0.174&0.0764 \\
5&{\rm +Auld}&245.35/239&22.88/17\ \ &50.76\ &-0.085&0.175&0.0770 \\
6&&242.43/239&23.46/17\ \ &46.32\ &-0.086&0.175&0.0766 \\
7&{\rm +Ritchie}&276.50/267&28.21/20\ \ &50.79\ &-0.086&0.175&0.0766 \\   
8&&273.59/267&27.94/20\ \ &45.36\ &-0.086&0.174&0.0764 \\   
9&{\rm +Wiedner}&317.40/306&35.12/22\ \ &49.86\ &-0.086&0.176&0.0768 \\
10&&315.21/306&33.12/22\ \ &45.53\ &-0.087&0.175&0.0768 \\
11&{\rm +Joram\ I}&431.89/386&37.97/26\ \ &48.98\ &-0.083&0.174&0.0759 \\
12&&432.83/386&35.58/26\ \ &45.14\ &-0.083&0.173&0.0757 \\
13&{\rm +Joram\ II}&495.96/417&44.57/30\ \ &49.22\ &-0.084&0.174&0.0761\\
14&&500.67/417&42.64/30\ \ &45.99\ &-0.085&0.173&0.0761\\
\hline
15&{\rm Frank}&139.07/166&4.41/6\ \ \ &48.67\ &-0.083&0.174&0.0757 \\
16&&139.27/166&4.71/6\ \ \ &46.51\ &-0.083&0.174&0.0757 \\
17&{\rm PSI}&153.32/150&8.38/10\ \ &65.17\ &-0.095&0.180&0.0793\\
\hline
18&{\rm Bertin}+$\pi^-$&271.40/243&17.47/18\ \  &50.68 \ &-0.103&0.184&0.0826\\
19&&273.53/243&15.87/18\ \ &48.94\ &-0.105&0.184&0.0820\\
\end{tabular}

\caption{Results of the study. The fits were made including a contribution to
the total $\chi^2$ from Eq. \protect\ref{chisig}   of $\Delta E_{\Sigma}=1$.}
\label{results2}\end{table}

\begin{figure}[hbt]
\epsfysize=100mm
\epsffile{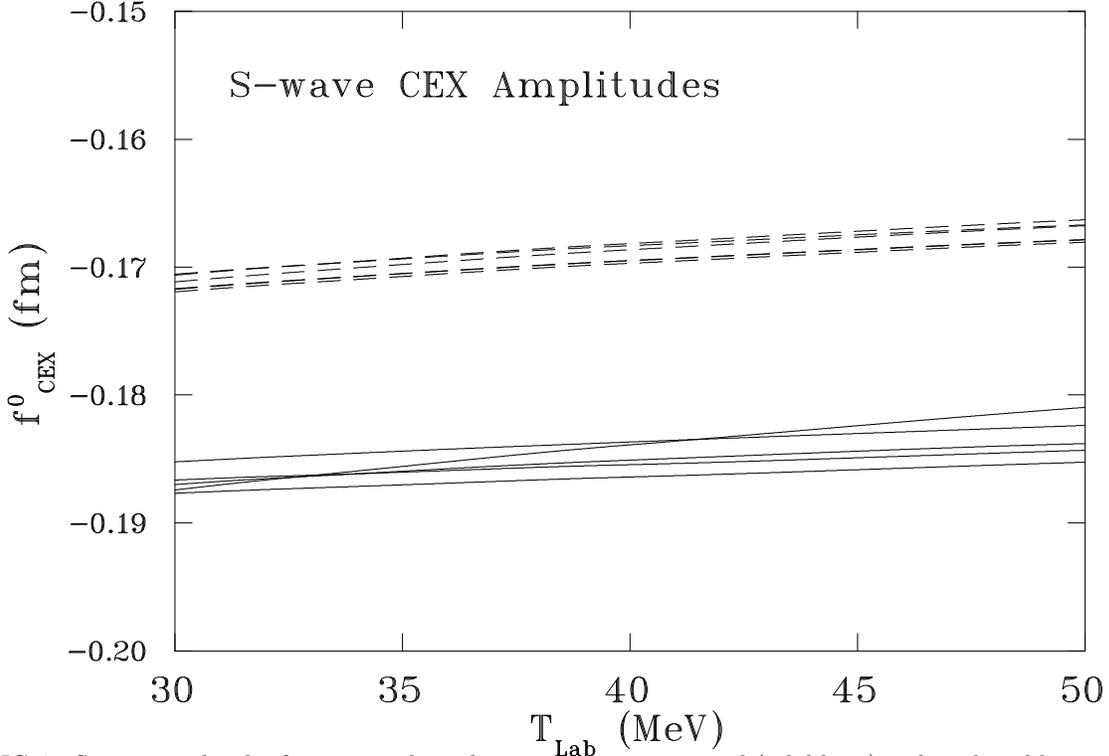}
\caption{S-wave amplitudes for pion-nucleon charge exchange, measured (solid
lines) and predicted by isospin from the present fit (dotted lines)}
\label{isospin} \end{figure}

\subsection{Phase Shifts and Low-energy Parameters}

Figures \ref{swaves}, \ref{p31p13}, \ref{p11} and \ref{p33} show the phase
shifts resulting from the fit 14 of Table \ref{results2}.  

We obtain for the isoscalar and isovector scattering lengths:

\begin{equation} 
b_0=0.001\pm 0.003\ \mu^{-1},\ \ {\rm and}\ \ b_1=-0.086\pm 0.003\ 
\mu^{-1}, 
\end{equation}
in reasonable agreement with the values obtained by Sigg et al.\cite{sigg}.

A recent QCD calculation by Bernard et al.\cite{bernard} gives $-0.096 
\mu^{-1}\le
b_1 \le -0.088 \mu^{-1}$  with which we are in marginal agreement.
 
The Chew-Low theory\cite{henley}  predicts the values of the scattering volumes
for all of the p-waves.  For the P$_{33}$ the prediction is $0.191 \mu^{-3}$
(using $f^2=0.0765$), in reasonable agreement with the values obtained.

An interesting feature of the P$_{31}$ and P$_{13}$ partial wave phase shifts
(Figure \ref{p31p13}) is that they are very nearly equal at low energies, 
becoming essentially identical at threshold.  Above 20 MeV in kinetic energy
the curves separate.   The Chew-Low theory predicts that these partial  waves
are always equal.  Improved models\cite{ew,oset2} give different magnitudes for
these scattering volumes, but the symmetry of the hamiltonian requires that
they be equal.  

It is not clear how fundamental this symmetry should be considered to be, and
thus on what level we should expect it to hold.  It has been known from some
time that this spin-isospin symmetry holds in the Skyrmion model\cite{mattis}.
Recently it has been shown\cite{dashen} that the relations among partial waves
of the Skyrmion model hold more generally in the large $1/N_c$ expansion  of
QCD. Since the predictions are for infinitely massive nucleons the symmetry
might well be expected to be broken above threshold where recoil effects become
important.

Comparisons with data have always shown this symmetry {\it not} to be
satisfied in nature, but we find that it {\it is} satisfied within the errors. 
We find for both scattering volumes $-0.028\pm 0.003 \mu^{-3}$.

Figure \ref{p31p13} shows that the KH80 phase shifts \cite{kp} follow  our
curve for values above 50 MeV but deviate at 20, 30, 40 and 50 MeV, just at the
Bertin energies.  Table \ref{results3} shows that, indeed, the analysis using
the Bertin data results in a $P_{31}$ scattering volume considerably larger
than the $P_{13}$.

For the P$_{33}$ phase shift (Fig. \ref{p33}) we note that the data fit  in the
region of 30 MeV to 86 MeV give the behavior of this resonant phase shift with
a reasonable accuracy.  

The P$_{11}$ phase shift (Fig. \ref{p11}) is seen to have a negative  excursion
at low energies and to cross zero around 140 MeV, as has been typical in
previous fits. However, in the present case, it is the low-energy data alone
which find this behavior, rather remarkable in view of the small size of the
amplitude.

\begin{table}[htb]\begin{tabular}{rlrrrrrrrr}
&&{\rm b$_3$}&{\rm b$_1$}&{\rm P$_{33}$}&{\rm P$_{13}$}&{\rm P$_{31}$}
&{\rm P$_{11}$} \\
1&{\rm Brack }&-0.050&-0.068&0.181&-0.031&-0.029&-0.102 \\
2&&-0.049&-0.064&0.178&-0.031&-0.031&-0.098 \\
3&{\rm +Frank }&-0.050&-0.068&0.179&-0.027&-0.028&-0.104 \\
4&&-0.049&-0.065&0.176&-0.027&-0.029&-0.102 \\
5&{\rm +Auld}&-0.050&-0.068&0.179&-0.027&-0.027&-0.105 \\
6&&-0.049&-0.065&0.176&-0.027&-0.028&-0.102 \\
7&{\rm +Ritchie}&-0.050&-0.068&0.180&-0.027&-0.026&-0.106\\
8&&-0.048&-0.064&0.177&-0.028&-0.028&-0.102 \\
9&{\rm +Wiedner}&-0.050&-0.073&0.178&-0.026&-0.029&-0.108\\
10&&-0.048&-0.064&0.177&-0.027&-0.028&-0.102\\
11&{\rm +Joram\ I}&-0.049&-0.071&0.179&-0.027&-0.028&-0.108\\
12&&-0.048&-0.063&0.177&-0.028&-0.028&-0.103\\
13&{\rm +Joram\ II}&-0.049&-0.071&0.178&-0.025&-0.028&-0.106\\
14&&-0.049&-0.064&0.176&-0.026&-0.027&-0.102\\
\hline
15&{\rm Frank}&-0.049&-0.071&0.169&-0.015&-0.025&-0.109 \\
16&&-0.049&-0.065&0.168&-0.014&-0.025&-0.108\\
17&{\rm PSI}&-0.055&-0.083&0.183&-0.027&+0.392&-0.119\\
\hline
18&{\rm Bertin}+$\pi^-$&-0.049&-0.076&0.178&-0.023&-0.033&-0.100\\
19&&-0.049&-0.070&0.177&-0.024&-0.034&-0.094\\
\end{tabular}
\label{results3}
\caption{Effective ranges (in units $\mu^{-1}$) for the s waves 
(b$_3$ and b$_1$) and scattering volumes (in units $\mu^{-3}$) 
from the fit.  The entries in this table correspond to those in Table 
\protect\ref{results2}.}\end{table} 

\subsection{Isospin Breaking}

We have calculated the prediction for the charge-exchange amplitude for each of
the fits shown in Table \ref{results2} in the same manner as discussed in  Ref.
\cite{gak}.  

The charge-exchange data fit are the same as before
\cite{sadler,salomon,bagheri}.  Note that the fit to this data is direct, it
requires no model and could have been done with a simple polynomial.  The
normalization of the data is constrained with the use of the TRIUMF data
\cite{salomon,bagheri}.  While these data  provide only limited angular
information, they give a relatively accurate value for the integrated
charge-exchange cross section and thus provide an important normalization
constraint.

While in the previous work\cite{gak} we demonstrated a small dependence on the
model used for the charged pion scattering, here we show (Fig. \ref{isospin}) a
comparison of the charge-exchange amplitudes determined directly from the
measurements  (the same as in \cite{gak}) with the predictions of six of the
fits from Table \ref{results2} (1, 2, 11, 12, 15 and 16).  As can be seen, the
variations among the data sets is small compared with the discrepancy.

It is worthwhile to discuss the size of the result.  It is possible to obtain a
large {\it fraction} for the breaking because of the smallness of the basic
amplitude itself. Let us compare with the NN case. If we assume that the {\it
potential} causing the breaking is the same for $\pi$N and NN  (as happens to
be the case for $\rho\omega$ mixing\cite{birbriar}) then  (in Born 
approximation) the breaking amplitude should be smaller for the pion-nucleon
case by a factor of the ratio of the pion mass to the nucleon mass (about
0.14).  However, a typical pion-nucleon amplitude at low energy (0.1-0.2 fm) is
a factor of 100 smaller than  a typical nucleon-nucleon amplitude (15-20 fm). 
Hence the percentage  of the breaking should be around 10 times larger in the
pion case  compared to the nucleon-nucleon case. The breaking amplitude
observed here  ($\sim$ 0.012 fm) would correspond to a breaking in NN case of
around 0.1 fm  which is too small to be observed at the present time.

The predictions with the Bertin data, from line 19 in Table \ref{results2} 
(not shown in the figure) agree with the measured charge exchange 
(as previously observed \cite{gak}).

A possible explanation has been  advanced by Piekarewicz \cite{piekarewicz} for
this breaking in terms of the difference of $\pi^0$ pion-nucleon coupling
constants for the neutron and proton.  This explanation would put the breaking
entirely in the charge-exchange channel or the $a_7$ coefficient 
in the notation of Ref. \cite{kg}. 

\subsection{The Pion-nucleon Coupling Constant}

The GMO (Goldberger-Miyazawa-Oehme) \cite{gmo} sum rule provides a useful
relation \cite{workman} between the isovector combination of the $\pi N$
scattering lengths and the $\pi NN$ coupling constant $f^2$. This sum rule
is obtained from a forward dispersion relation for the invariant amplitude 
C$^{-}$ subtracted at $\omega_{LAB}=0$ and evaluated at 
$\omega_{LAB}=\mu $ \cite{hoehler}.
 
\begin{equation} 
4\pi \frac{m+\mu }{3m\mu }
(a_1-a_3)=\frac{8\pi f^2}{\mu ^2-\omega _B^2}+4\pi J, 
\end{equation}
where 

\begin{equation} 
J=\frac1{2\pi ^2}\int\nolimits_0^\infty 
\frac{\sigma ^{-}(k)}{\omega (k)}dk.
\end{equation}

In this expression $k$ is the incident pion laboratory momentum and $\omega
_B=-\mu ^2/2m.$  The isospin-odd total cross section is defined by $\sigma
^{-}=\frac 12(\sigma _{-}-\sigma _{+})$ where $\sigma _{\pm }=\sigma _T(\pi
^{\pm }p).$ 

To evaluate J we have used the SM95 \cite{sm95} phase shift analysis for $k <
2$ GeV/c, our own parameterization for 2 GeV/c $< k < 4$ GeV/c (a fit to total
cross section data taken from the Review of Particle Properties \cite{rpp}),
and a Regge fit for 4 GeV/c $< k < $ 240 GeV/c taken from the same source. The
integral was truncated above 240 GeV/c. The resulting value of the integral is
J=(--1.308+0.068+0.157) mb =-1.081 $\pm$ 0.005 mb where the contributions from
each momentum interval is shown. If we were to assume that the Regge fit is
valid to infinite momenta, the contribution to J coming from above 240 GeV/c is
0.030 mb, which would lead to J=-1.051 mb. For the following discussion we will
use the truncated value, J=--1.081 mb.  This value  can be compared with
J=--1.072 mb obtained by Ref. \cite{workman} and  J=--1.077 mb quoted in Ref.
\cite{workman} from an unpublished preprint by R. Koch. Thus it would appear
that uncertainty in the determination  of J is the order of 1/2 to 1\%.  Since
it contributes about 1/3 of the value of $f^2$, the error in $f^2$ from this
source is less than 1/3 \%. Locher and Sainio\cite{ls} concluded, however, that
the uncertainty in  J was slightly larger, leading to a 1\% error in $f^2$.

Thus, we have the simple formula for $f^2$ ($a_1$ and $a_3$ in pion mass 
units) 
\begin{equation} 
f^2=0.0269+0.1904(a_1-a_3). 
\end{equation} 

Some of the results of our fits to the  low-energy $\pi ^{\pm }p$ elastic
scattering data are shown in Tables  \ref{results1} and \ref{results2}. The
resulting average value of the coupling constant is $f^2=0.0764\pm 0.0007$ 
where the error quoted includes our fitting error only. 

The values for the scattering lengths advocated by SM95, $a_3=-0.087\,\mu
^{-1}$ and $a_1=0.175\,\mu ^{-1}$, lead to $f^2=0.0768$ when used in our
relation. Arndt et al. \cite{arndt}  quote a value of $f^2=0.076\pm 0.001$ 
while Markopoulou-Kalamara and Bugg \cite{bugg} found $f^2=0.0771\pm 0.0014$.
Timmermans quotes a preliminary value from his $\pi$N 
analysis\cite{timmermans} of $f^2=0.0741\pm 0.0008$ (statistical error only).

Thus recent analyses of the charged pion coupling constant from pion-nucleon
scattering seem to be in moderately good agreement.  It is  very interesting to
know if this value is consistent with that obtained from the  nucleon-nucleon
interaction since that comparison serves as a check on  our theory of the
strong interaction.

The Nijmegen group \cite{nijmegen} found $f^2=0.0748\pm 0.0003$ for the charged
pion coupling constant and $f^2=0.0745\pm 0.0006$\ for the neutral pion.
Ericson et al.\cite{loiseau} find from the analysis of np charge-exchange 
data, $f^2=0.0808\pm 0.0017$.

While results here would seem to confirm (perhaps even with smaller  errors due
to the determination of the scattering lengths directly  from the low-energy
data) the results of other analyses of pion-nucleon scattering, there is an
important caveat which should be mentioned in regard to all of the analyses
with the GMO sum rule.  It is the isovector scattering length which enters in
the GMO relation.  As we saw in the previous section, there is a strong
indication of an isospin breaking  from the comparison of the elastic
scattering determination of this amplitude with its determination from charge
exchange.  If the explanation advanced in Ref. \cite{piekarewicz} is correct
(so that all of the breaking is in the  charge-exchange channel) then the GMO
relation will lead to a  correct value of $f^2$.  If there is  some breaking in
the elastic scattering channels then that correction should be made before
applying the  GMO  relation.  In the extreme case that all of the  breaking
comes in the elastic amplitudes (an amplitude of the form of  $a_3$\cite{kg},
possibly from $\rho -\omega$ mixing, with the opposite sign than predicted) and
the isovector amplitude from charge exchange is the correct one then the
pion-nucleon coupling constant would return to the ``large'' values.  It is not
possible to reliably determine $f^2$ (from GMO) until the question of the
origin of the  isospin breaking is resolved.

\subsection{The $\Sigma$ Term}

\begin{figure}[htb]
\epsfysize=100mm
\epsffile{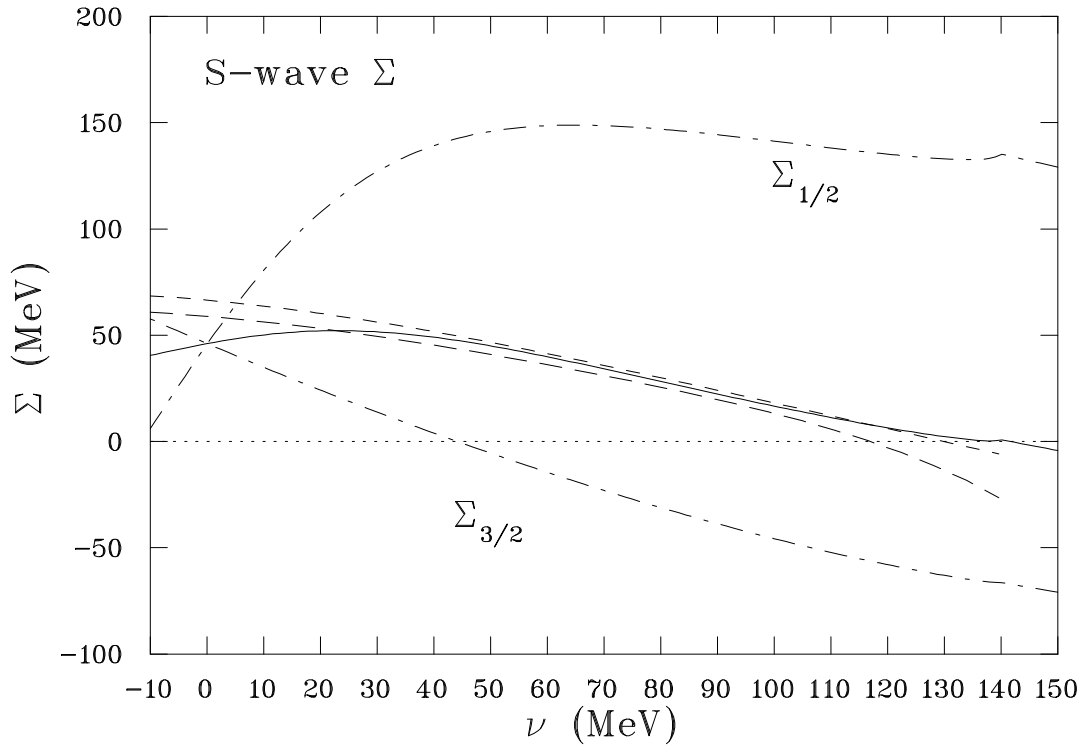}
\caption{S-wave $\Sigma$ term amplitude vs $\nu$ from the present work 
(solid curve) 
compared with the pole-subtracted expansions of 
Nielsen and Oades\protect\cite{no} (short dash) and H\"ohler
\protect\cite{hoehler4} (long dash). The two isospin components from our
work are shown with the dash-dot curves.}
\label{sigma2} \end{figure}

Figure \ref{sigma2} shows the amplitude defined by Eq. \ref{sigdef} as a 
function of $\nu$ (which is roughly equal to the center of mass pion energy,
$\omega$, in this region). Since above threshold the amplitude, the real part
was used.  The result of the solid curve at $\nu=0$ represents the s-wave
contribution to $\Sigma$ in our model. The solid and dash-dot curves were
calculated  using the Jost solutions for fit 14 in Table \ref{results2}. The 
kinematic singularity at threshold is clearly visible. 

While this  plot is given as a function of $\nu$, the more relevant variable
for  extrapolation may be the center of mass momentum.  Since the CD point lies
at  $k\approx i\mu$, a circle in the complex plane with radius $\mu$\ passes
through the real axis at a kinetic energy around 57 MeV, or in the center of
range studied in this work.

The values obtained for $\Sigma$ lie around 48$\pm 4$ MeV.  The error  quoted
is determined from the variation among data sets and does not include the model
error. It is interesting to note that there is very little difference between
the results from the new $\pi^+$\  data and that of Bertin et al.
\cite{bertin}. This is perhaps understandable from the observation that the
Bertin data for the cross section are 20\% higher than the prediction from the
fits (see Fig. 1) so the $\pi^+$ p amplitude can be expected to be  10\%
higher.   Since the $\pi^+$\ and $\pi^-$\ contribute equally  to $\Sigma$, the
difference of the analysis between the two data groups can be expected to be of
the order of 5\% or 2-3 MeV.  Indeed the difference observed is of that order. 

While the usually accepted value of $\Sigma$ is around 65 MeV \cite{sainio}, 
this is not the first time that a smaller value has been obtained. Ericson 
found a similar value ($44\pm 6$ MeV) \cite{ericson}. The modern approach 
\cite{sainio,hoehler2} seems to be to consider this value as not including the
part of the $\Sigma$ term coming from the cut in $t$.  This additional part
(about 11 MeV) is added on after the extrapolation of the pion-nucleon
amplitude. If we compare our value of $\Sigma$ with $\Sigma_d$ of Ref. 
\cite{sainio}  (before correction for the $\pi\pi$ channel) we are in good
agreement. However, the $\pi\pi$ channel is implicitly included in the present
analysis since an integral over the discontinuity of the $t$ cut gives the
potential.  Thus the singularities in $t$ are present in the model and we must
consider the value obtained as  including this correction.

\subsection{Off-shell Amplitudes\label{osa}}

\begin{figure}[htb]
\epsfysize=100mm
\epsffile{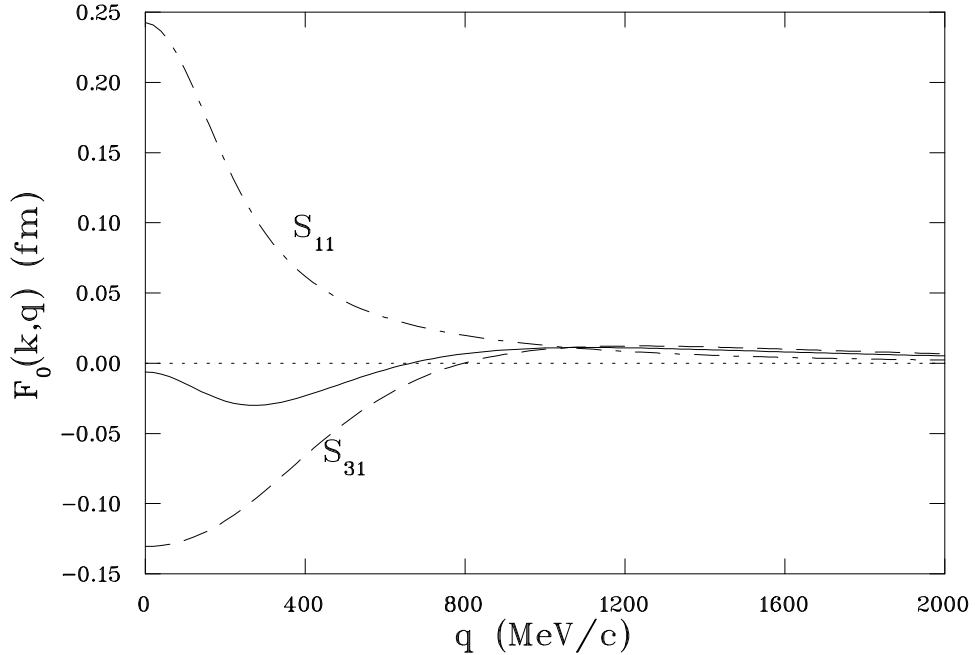}
\caption{S-wave off-shell amplitudes. The solid line corresponds to the
isoscalar combination of the two amplitudes.}\label{offs} \end{figure}

\begin{figure}[htb]
\epsfysize=100mm
\epsffile{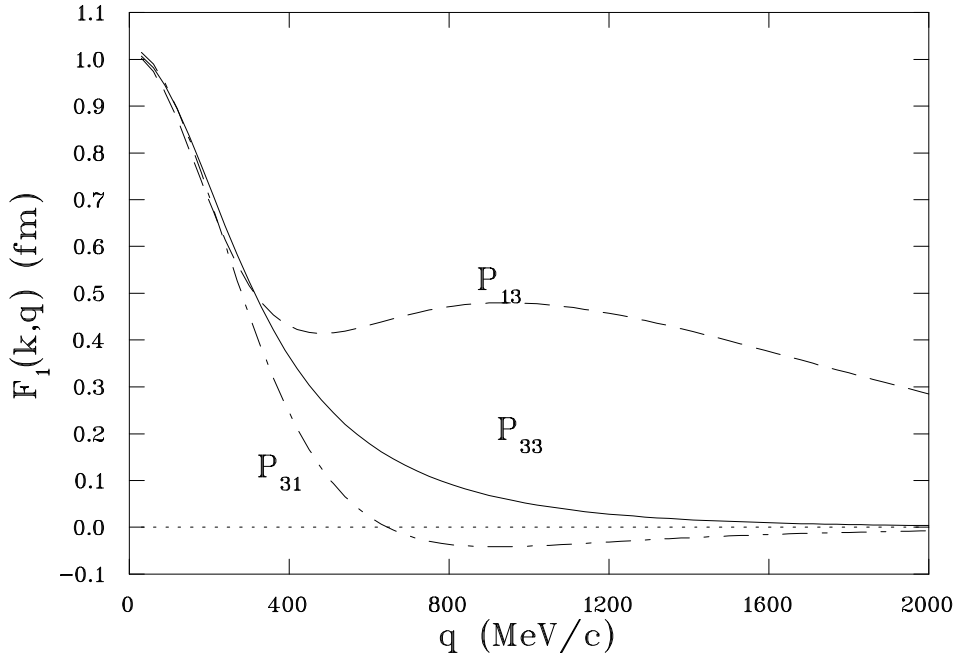}
\caption{P-wave off-shell amplitudes. Plotted is the ratio
of the off-shell value to the on-shell value at 5 MeV.}\label{offp} 
\end{figure}

A knowledge of the off-shell amplitude is necessary for the treatment  of
pion-nucleus scattering. An early determination of these amplitudes  was made
by Landau and Tabakin \cite{lt} using the assumption of  a separable potential
and deducing the form of the off-shell  amplitude from the energy dependence of
the on-shell amplitude.  Since the knowledge of the scattering amplitude was
needed in the region where the scattering becomes inelastic, the potentials
obtained  were complex, not a very satisfactory situation.  Londergan, McVoy
and Moniz \cite{lmm} were able to  find real functions for the off-shell
dependence by considering only the potentials in the elastic channel, including
the inelasticity by coupling to one other 2-body channel.

In our case, since we are fitting to data only in the elastic region, there is
no problem of a complex extension.  We can calculate the off-shell amplitudes
directly by the Jost formula given in Section \ref{josts} for the s waves  or by
direct numerical solution for the p waves.   Figures \ref{offs} and  \ref{offp}
show the ratio of the off-shell amplitudes to the value on-shell at 5 MeV.

Of particular interest is the isoscalar combination of the s-wave 
amplitudes.  At low energy this combination nearly vanishes on shell,
but this cancellation does not necessarily occur off shell. As can be seen 
from Figure \ref{offs}, for larger values of momentum there is no 
cancelation at all.

Recent measurements\cite{meyer} have resulted in accurate threshold cross
sections for $\pi^0$ production in pp collisions.  The most common calculation
of pion production achieves the needed momentum sharing among the nucleons  by
rescattering of the meson after it is emitted from one of the protons.  Since
the s-wave $\pi^0$N  scattering length is near zero,  this contribution would
seem to be small at threshold.  Based on this assumption, Horowitz et al.
\cite{horowitz}  and Lee and Riska \cite{lee} constructed models based on 
heavy-meson exchange which were able to explain the data.  

Using the fact that the $\pi^0$ is produced ``off-shell,'' Hern\'andez and Oset 
\cite{oset} were able to explain the cross section using an estimated
dependence of the off-shell behavior.  Their results are uncertain, however,
due to a lack of knowledge of this dependence.

For the p waves (Fig. \ref{offp}) it can be seen that the off-shell  dependence
of the P$_{31}$ and P$_{13}$ (as well as the P$_{33}$)  amplitudes is nearly
identical below 300 MeV/c.  This fact indicates  that the long-range part of
the interaction is the same for these three  waves.

\subsection{Partial Total Cross Sections}

Recently transmission measurements have been made \cite{freidman,kriss} to
determine the integrated elastic cross section beyond a fixed angle: ``partial
total cross sections''.   While many of these  measurements were made   at
higher energies than those treated here, there are several points in the energy
region of interest.  We have not included these points in the fit but now
compare with the values obtained.  Until recently the two measurements have
differed but now there seems to be general agreement between them, but
disagreement with the Brack data above 50 MeV.   Our values are shown in Table
\ref{pts}.  

\begin{table} \begin{tabular}{lcccc} T$_{\pi}$\ {\rm (MeV)}&
{\rm Present\ Work}& {\rm Ref \cite{freidman}} &{\rm Ref\cite{kriss}}\\
39.8&7.6&&8.5$\pm$0.7\\ 44.7&8.9&&9.2$\pm$0.8\\ 45.0& 9.0&10.8$\pm$ 1.0\\
51.7&11.3&&11.8$\pm$0.8\\ 52.1& 11.5&12.4$\pm$ 1.0\\ 54.8&12.6&&13.2$\pm$ 0.5\\
59.3&14.7&&15.8$\pm$ 0.4\\ 63.1& 16.6&18.0$\pm$ 0.6\\ 66.3&18.4&&20.4$\pm$ 0.4\\
67.5&19.2&20.7$\pm$ 0.6\\ 71.5&21.7&23.8$\pm$ 0.6\\ 80.0&28.1&&29.7$\pm$ 0.7\\
92.5&39.6&43.3$\pm$ 1.5\\ \end{tabular}
\caption{Partial total cross sections in
mb at the 30$^{\circ}$ limit.} \label{pts}\end{table}

\subsection{Polarization Asymmetry}

\begin{figure}[htb]
\epsfysize=100mm
\epsffile{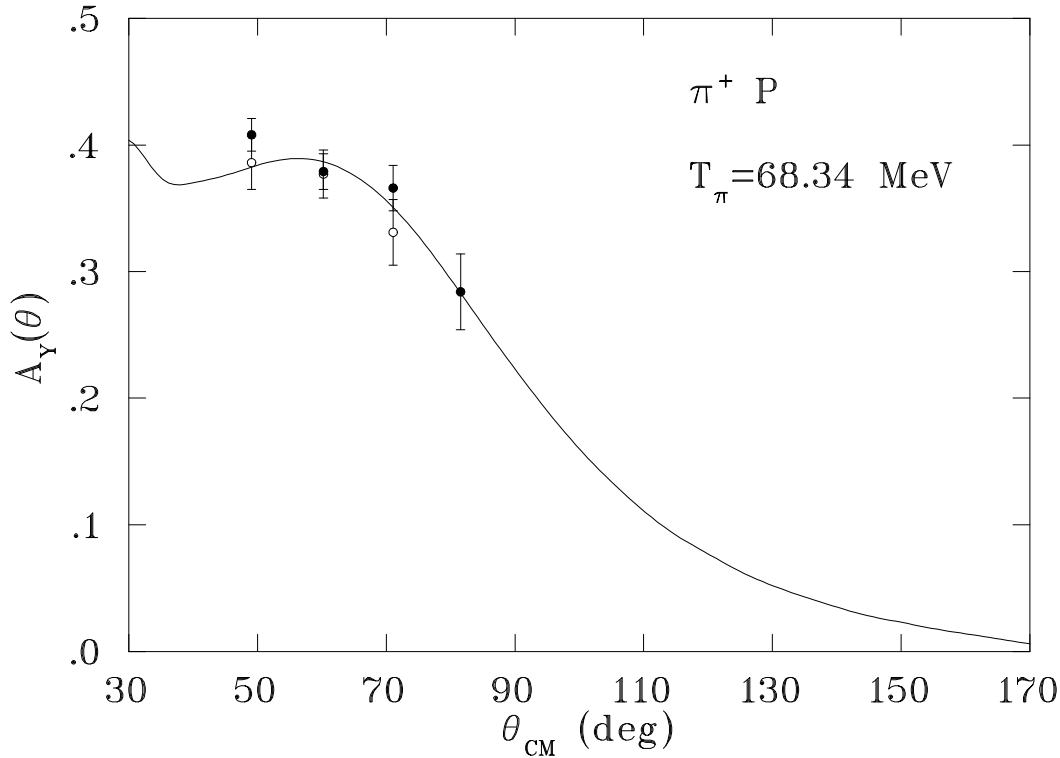}

\caption{Comparison of the prediction of the fit with the polarization
asymmetry data of Wieser et al. \protect\cite{wieser1,wieser2}}\label{ay} 
\end{figure}

While there is no published data on the polarization asymmetry in low-energy
charged pion scattering there is one measurement at 68.34 MeV  which has
appeared in conference proceedings \cite{wieser1} and in a thesis
\cite{wieser2}.  The prediction of our fits are in excellent agreement with 
these data as shown in Fig. \ref{ay}.

\subsection{Gaussian Fits\label{gauss}}

While a Gaussian potential has little theoretical justification, some fits 
were made with this form to test the sensitivity to the model potential.  If a
result is stable under this change it may well be believed to be  weakly
dependent of the form of the potential function.  The results are shown in
Table \ref{resultgss}.  It is seen that the values for the pion-nucleon
coupling constant remain within a very narrow range, the same range as for the
exponential potential, and thus are not sensitive to  the potential form
chosen.

{\scriptsize 
\begin{table}[htb]
\begin{tabular}{rlrrrrrrrrrrr}
&&$\chi^2$/{\rm N\ (Data)}&$\chi^2$/{\rm N\ (N)}& $\Sigma$ {\rm (MeV)}
&$a_3$\ ($\mu^{-1}$)&$a_1$\ ($\mu^{-1}$) &$f^2$\ \ &P$_{33}$\ &P$_{13}$\ 
&P$_{31}$\ &P$_{11}$\ \\
1&{\rm Brack }&68.25/62\ &13.45/10\ \ 
&22.59\ &-0.085\ &0.174\ &0.0762&0.168 &-0.029&-0.028&-0.061 \\
2&{\rm +Frank }&231.72/228&26.64/16\ \ 
&22.86\ &-0.089\ &0.175\ &0.0771&0.166&-0.026&-0.026&-0.063\\
3&{\rm +Auld}&242.71/239&27.48/17\ \ 
&22.81\ &-0.089\ &0.175\ &0.0771&0.166&-0.026&-0.026&-0.063\\
4&{\rm +Ritchie}&272.89/267&31.90/20\ \ 
&22.83\ &-0.089\ &0.175\ &0.0771&0.167&-0.026&-0.025&-0.063\\
5&{\rm +Wiedner}&314.50/306&37.36/22\ \ 
&22.89\ &-0.089\ &0.175\ &0.0771&0.166&-0.026&-0.025&-0.063\\
6&{\rm +Joram\ I}&441.25/386&39.86/26\ \ 
&22.86\ &-0.086\ &0.173\ &0.0762&0.167&-0.027&-0.025&-0.064\\
7&{\rm +Joram\ II}&516.30/417&46.08/30\ \ 
&23.14\ &-0.087\ &0.173\ &0.0764&0.166&-0.025&-0.025&-0.064\\
\end{tabular}

\caption{Results of the study with a Gaussian potential with the condition
that $\Delta E_{\Sigma}=1$\ MeV.}
\label{resultgss}\end{table} }

Another result which is only weakly dependent of the potential form are the
scattering volumes of the $P_{31}$ and $P_{13}$ partial waves.  They are
again found to be equal, albeit slightly smaller than for the exponential
case.

The values of other scattering volumes are altered significantly, especially
the $P_{11}$ wave which is small in this region and hence difficult to
determine.

The value of the $\Sigma$ term is found to vary by less than $\pm$ 1/2 MeV over
the data sets, but  with a different value than for the exponential potential. 
Thus, in the KG model, the extrapolation to the CD point depends  on the form
of the assumed potential. 

\newpage

\section{Summary}
We have analysed recent low-energy pion-proton elastic scattering data.
The principal findings are:
\begin{enumerate}
\item The large bulk of the modern elastic scattering data is internally
consistent and inconsistent with the older ``Bertin'' data.

\item The observation of the isospin violation previously indicated was
confirmed for the variations in the data set fit over a larger range in
energy.

\item The value of the subthreshold parameter, the $\Sigma$ term, has been
extracted with the exponential potential and the result is a value around 
50 MeV, lower than previous estimates.  The smallness of this value was 
shown not to be because of the change in data (which had little effect) 
but due to the model which was used. This conclusion is reinforced by the
study with the Gaussian potential (considered unrealistic) in which a 
yet smaller value was found.

\item The pion-nucleon coupling constant was extracted from the scattering
lengths with the use of the GMO sum rule and a value ($f^2=0.0764\pm 0.0007$) 
in agreement with most (but not all) modern determinations was obtained.

\item Scattering volumes were extracted for the p waves.  The parameters
for the P$_{13}$ and P$_{31}$ were found to be the same within errors, 
as predicted from the Chew-Low theory and various improvements \cite{ew}
and the limit of a large $N_c$ expansion of QCD\cite{dashen}.
The value of the P$_{33}$ scattering
volume was found to be smaller than previous determinations.
 
\item  Off-shell amplitudes for pion-nucleon scattering at low momentum
transfers were obtained.  The isospin-zero combination of the s-wave
amplitudes was shown to be relatively large off shell.  Rather remarkably,
the P$_{13}$, P$_{31}$ and P$_{33}$ have the same off-shell dependence for
momenta below 300 MeV/c.

\end{enumerate}

One of us (wrg) gratefully acknowledges discussions with T. E. O. Ericson, 
B. Loiseau, S. Coon and M. Mattis.

This work was supported by the U. S. Department of Energy.

\end{document}